\documentclass[12pt]{article}
\usepackage{graphicx}
\usepackage[utf8]{inputenc}
\usepackage[numbers,sort&compress]{natbib} 
\usepackage{hyperref}
\usepackage{amssymb}
\usepackage{subcaption}
\usepackage{setspace}

\newcommand\pubnumber{SNSN-323-63}
\newcommand\pubdate{\today}

\def\Title#1{\begin{center} {\Large #1 } \end{center}}
\def\Author#1{\begin{center}{ \sc #1} \end{center}}

\newcommand\pubblock{\rightline{\begin{tabular}{l} \pubnumber\\
         \pubdate  \end{tabular}}}
\newenvironment{Abstract}{\begin{quotation}  }{\end{quotation}}
\newenvironment{Presented}{\begin{quotation} \begin{center} 
             PRESENTED AT\end{center}\bigskip 
      \begin{center}\begin{large}}{\end{large}\end{center} \end{quotation}}
\def\Acknowledgements{\bigskip  \bigskip \begin{center} \begin{large}
             \bf ACKNOWLEDGEMENTS \end{large}\end{center}}

\def\beq{\begin{equation}}
\def\eeq#1{\label{#1}\end{equation}}
\def\eeqn{\end{equation}}

\def\beqa{\begin{eqnarray}}
\def\eeqa#1{\label{#1}\end{eqnarray}}
\def\eeqan{\end{eqnarray}}

\let\bar=\overbar

\def\Dslash{\not{\hbox{\kern-4pt $D$}}}
\def\dslash{\not{\hbox{\kern-2pt $\del$}}}

\def\msb{{\bar{\ssstyle M \kern -1pt S}}}

\renewcommand*{\thefootnote}{\fnsymbol{footnote}}

\begin{document}
\begin{titlepage}
\pubblock

\vfill
\Title{Measuring the CP structure of the top Yukawa coupling in $t\bar{t}H$ events at the LHC}
\vfill
\Author{Emanuel Gouveia$^1$
\\[2mm]
{\rm also on behalf of}
\\[2mm]
Duarte Azevedo$^{1,2}$,
Ricardo Gon\c{c}alo$^3$,
António Onofre$^4$
\\[3mm]
{\footnotesize {\it 
$^1$ LIP, Departamento de F\'{\i}sica, Universidade do Minho, 4710-057 Braga, Portugal\\
$^2$ Centro de F\'{\i}sica Te\'{o}rica e Computacional, Faculdade de Ci\^{e}ncias, Universidade de Lisboa, Campo Grande, Edif\'{\i}cio C8 1749-016 Lisboa, Portugal\\
$^3$ LIP, Av. Prof. Gama Pinto, n.2, 1649-003 Lisboa, Portugal;\\ Faculdade de Ci\^{e}ncias da Universidade de Lisboa, Campo Grande, 1749-016 Lisboa, Portugal\\
$^4$ Departamento de F\'{\i}sica, Universidade do Minho, 4710-057 Braga, Portugal\\
}
}
}
\vfill
\begin{Abstract}

The ATLAS and CMS collaborations recently announced the observation of the associated production of the Higgs boson with a top quark pair ($t\bar tH$) at the LHC. This process depends directly on the the top quark Yukawa coupling and provides access to its properties. In particular, a CP-odd component is allowed in models beyond the Standard Model with extended Higgs sectors.

Studies of the feasibility of such a measurement at the 13 TeV LHC were carried out, in the $H\rightarrow b\bar b$ decay channel and with the $t\bar t$ system decaying through the semileptonic and dileptonic channels. Fast detector simulation and kinematic fits were applied to samples of SM backgrounds and of signal scenarios with different CP-mixing angles of the coupling. Ratios of projections of momenta and angular distributions using boosted reference frames were found to be sensitive to the CP-mixing angle. Those are expected to be robust relatively to modeling uncertainties and are thus presented as good candidates for experimental use.

Expected confidence levels for the exclusion of scenarios with a CP-odd component in the top quark Yukawa coupling were obtained, using different observables as discriminants, and are presented for integrated luminosities up to 3 ab$^{-1}$, as expected after the full HL-LHC program.

\end{Abstract}
\vfill
\begin{Presented}
$11^\mathrm{th}$ International Workshop on Top Quark Physics\\
Bad Neuenahr, Germany, September 16--21, 2018
\end{Presented}
\vfill
\end{titlepage}
\def\thefootnote{\fnsymbol{footnote}}
\setcounter{footnote}{0}

\section{Introduction}

The production of a Higgs boson in association with a top quark pair ($t\bar tH$) at the LHC has been recently observed by the ATLAS~\cite{Aaboud:2018urx} and CMS~\cite{Sirunyan:2018hoz} collaborations, providing direct proof of the top quark-Higgs interaction. In models beyond the Standard Model (BSM), the top quark Yukawa coupling governing this interaction may have a CP-odd component~\cite{Fontes:2017zfn} and the corresponding Lagrangian can be parametrised as
\begin{equation}
 \mathcal{L}=\kappa y_t \bar t (\cos\alpha+i\gamma_5\sin\alpha)th.
\end{equation}
The Standard Model (SM) is recovered for $\alpha=0$ and $\kappa=1$ and a pure CP-odd scenario corresponds to $\alpha=\pm\pi/2$.

Measurements of the electron dipole moment give strong constraints on the CP-odd component of the top quark-Higgs interaction if one assumes a SM Yukawa coupling for the electron. However, if the electron Yukawa coupling also acquires a CP-odd component, this incompatibility can be mitigated~\cite{Fontes:2017zfn}. In future $e^+e^-$ colliders, a CP-violating top quark Yukawa coupling would lead to large tree level effects in $t\bar tH$ production, which may be measured with precision~\cite{BarShalom:1995jb}. An indirect constraint on the CP-odd component of the top quark Yukawa coupling can be obtained from the Higgs boson production at the LHC via gluon fusion. This process is loop-induced and, in the SM, dominated by the top quark loop. However, the translation between the production rate and the coupling requires assumptions about every other contribution to the loop, including those from heavy BSM particles. Assuming only SM contributions besides the top quark loop, the LHC Run I data still allows values of $\kappa\sin\alpha$ above 0.5~\cite{rohini2}.

Unlike Higgs production via gluon fusion, $t\bar tH$ production at the LHC depends directly on the top quark-Higgs interaction, which allows measuring the top quark Yukawa coupling without the need for the previously mentioned assumptions. For fixed $\kappa$, increasing the CP-odd component decreases the inclusive $t\bar t H$ production cross-section~\cite{rohini2}, but this is not sufficient to obtain $\kappa$ and $\alpha$ simultaneously. Several observables in $t\bar t H$ are sensitive to the CP nature of the top quark Yukawa coupling and could in principle be used to extract both parameters. One important feature in the kinematics of $t\bar t H$ with a CP-odd coupling is the higher fraction of boosted ($p_T\gtrsim200$~GeV) Higgs bosons with respect to the CP-even scenario~\cite{rohini2}. Regarding the top quark pair, a CP-odd coupling yields much larger separation in pseudorapidity and smaller separation in azimuth between the top quarks with respect to the CP-even scenario~\cite{demartin}. In the $t\bar t H$ rest frame, CP-even signal events often have one of the top quarks traveling in a direction very close to that of the Higgs boson, recoiling against the other top quark, while for CP-odd signal events the three directions are more evenly distributed~\cite{cpangvars}. Ratios of projections of top quark momenta were proposed as discriminants in \cite{gunion}, for example $b_4\equiv p^z_tp^z_{\bar t}/(|\vec{p_t}||\vec{p_{\bar t}}|)$. Spin correlation observables of the $t\bar t$ system in $t\bar tH$ also depend on the CP-odd component of the Yukawa coupling, with enhanced sensitivity in the boosted Higgs regime~\cite{Buckley:2015vsa}. Angular observables built from boosted reference frames were also shown to be sensitive to the $CP$ nature of the coupling~\cite{cpangvars}.

\section{CP-odd exclusion in the $H\rightarrow b\bar b$ channel}

Two analyses were carried out on the $H\rightarrow b\bar b$ final state of $t\bar tH$. The single-lepton analysis is presented in detail in \cite{duarte} and the dilepton analysis is presented in \cite{cpangvars}.

Events from $pp$ collisions were generated using {\sc MadGraph5\_aMC@NLO}~\cite{madgraph} at $\sqrt{s}=13$ TeV. Signal $t\bar t H$ events and the irreducible background $t\bar t b \bar b$ were generated at NLO in QCD.
Signal samples with a non-zero $CP$-odd component were generated with the \texttt{HC\_NLO\_X0} UFO model~\cite{demartin}.
Decays were handled by {\sc MadSpin}~\cite{madspin}. The SM backgrounds $t\bar t+$jets, $t\bar t+V$ ($V=Z,W$), single top, $V+$jets, $V+b\bar b$ and $VV$ were also generated.
Parton shower and hadronisation were performed with {\sc Pythia6}~\cite{pythia} and followed by fast detector simulation with the {\sc Delphes}~\cite{delphes} package.

Single-lepton events were selected with a number of jets between 6 and 8 and a number of $b$-tagged jets between 3 and 4. Dilepton events were selected with at least 4 jets, of which at least 3 must be $b$-tagged. Leptons and jets were required to have $p_T\geq20$ GeV and \mbox{$|\eta|\leq2.5$} in both analyses. In the single-lepton selection, missing transverse energy ($E_T^{\rm miss}$) was required to be above 20 GeV. In the dilepton channel, the dilepton mass $m_{\ell\ell}$ must pass the $Z$ mass veto, i.e., \mbox{$|m_{\ell\ell}-91\textrm{ GeV}|>10\textrm{ GeV}$}.

\begin{figure}[bh]
\begin{center}
\begin{tabular}{ccc}
\includegraphics[width=0.49\textwidth]{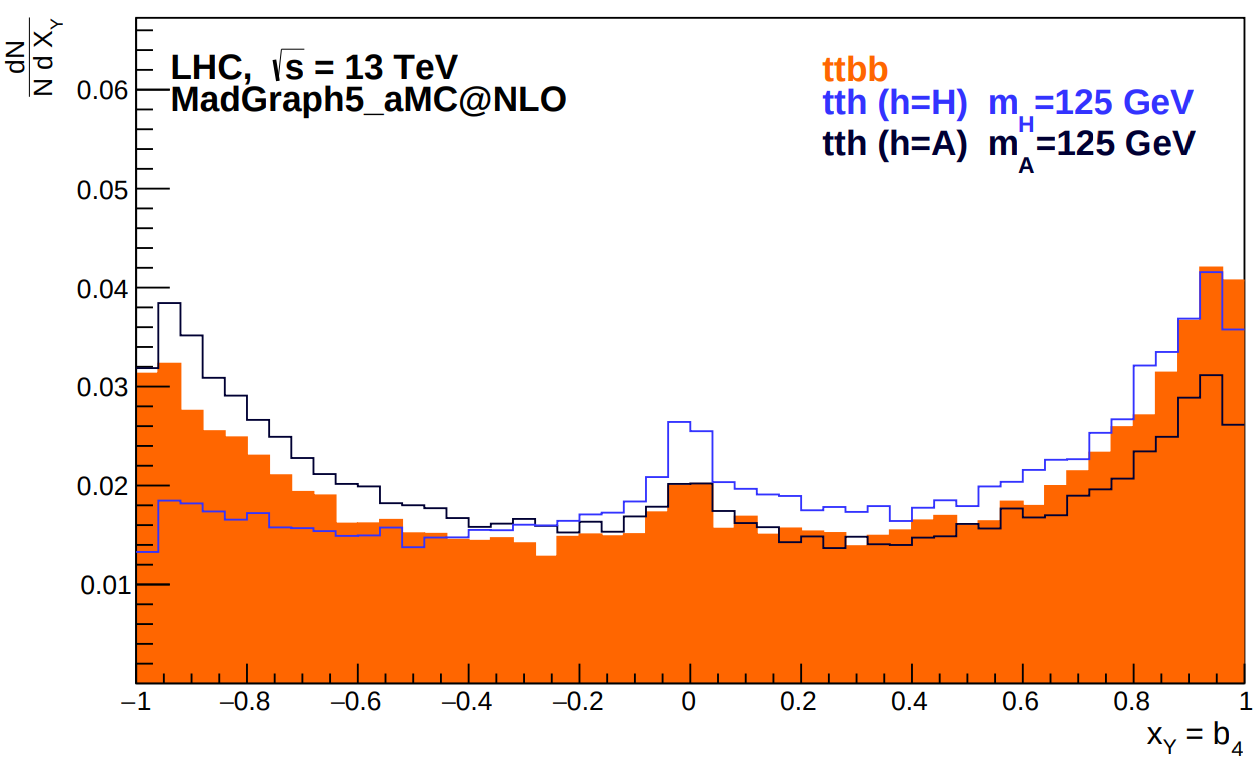} & \quad & 
\includegraphics[width=0.52\textwidth]{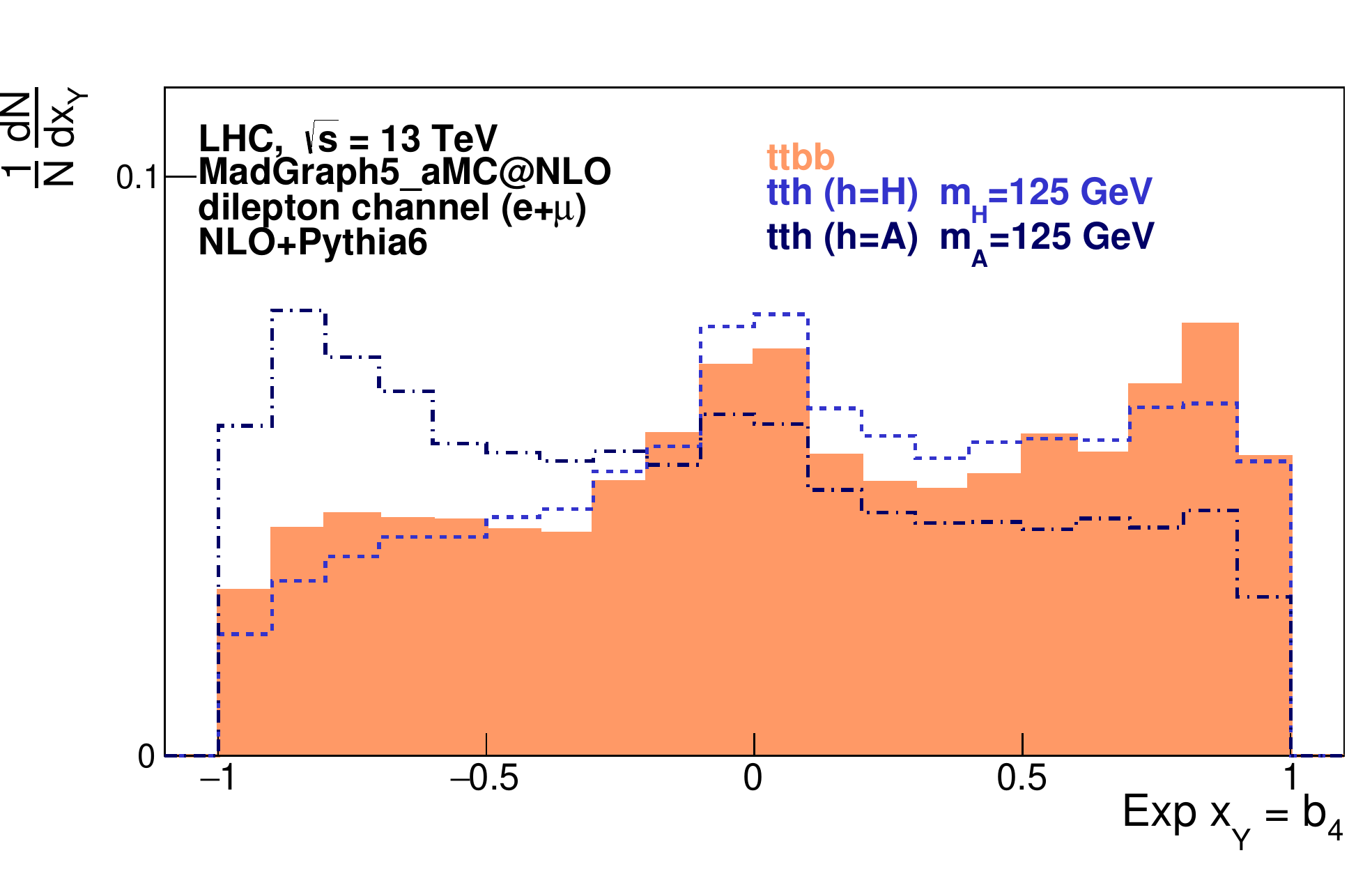}\\[-4mm]
\end{tabular}
\caption{Normalised distributions of $b_4$ for signal events in the CP-even ($t\bar t H$) and pure $CP$-odd ($t\bar t A$) scenarios and for the $t\bar t b \bar b$ background, after cuts and reconstruction. Left: single-lepton selection~\cite{duarte}; right: dilepton selection~\cite{cpangvars}.}
\label{dists}
\end{center}
\end{figure}

Kinematic reconstruction of events was performed in the single-lepton channel with the {\sc KLFitter} package~\cite{Erdmann:2013rxa}. In the dilepton channel, a boosted decision tree (BDT) was used to assign jets to the $b$-quarks from $t$, $\bar t$ and $H$.
A solution for the neutrino momenta was obtained by combining information from $E_T^{\rm miss}$ with a constraint on $W$ and $t$ masses.
Figure \ref{dists} shows normalised distributions of the $b_4$ observable after cuts and kinematic reconstruction for the SM signal ($t\bar t H$), pure $CP$-odd signal ($t\bar t A$) and irreducible background $t\bar t b\bar b$ events, separately for the single-lepton (left) and dilepton (right) channels.

\pagebreak

\begin{figure}[h]
\centering
\begin{subfigure}{0.49\textwidth}
 \includegraphics[width=\textwidth,clip]{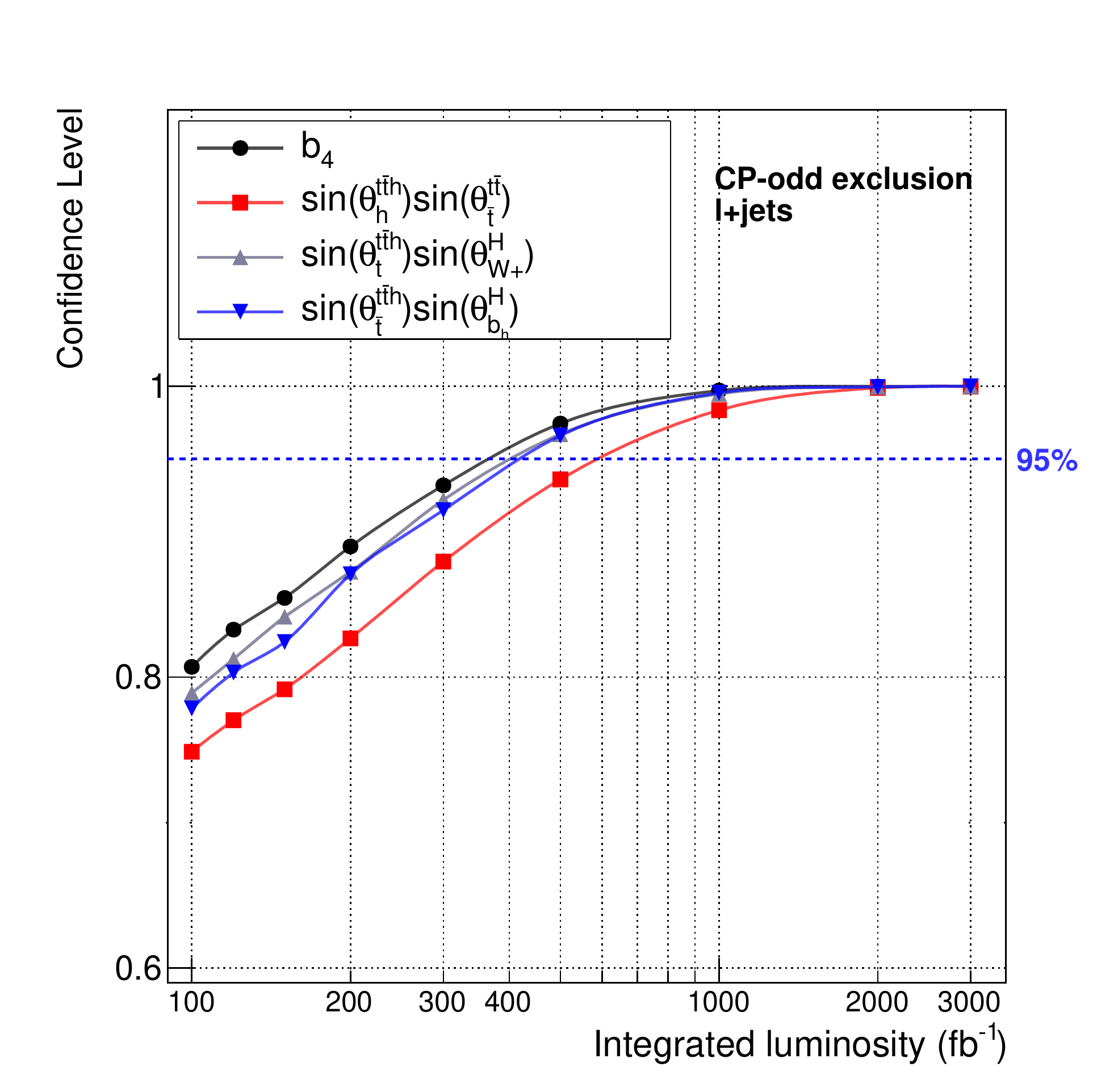}
 \caption{}
 \label{semilep}
\end{subfigure}
\hfill
\begin{subfigure}{0.49\textwidth}
 \includegraphics[width=\textwidth,clip]{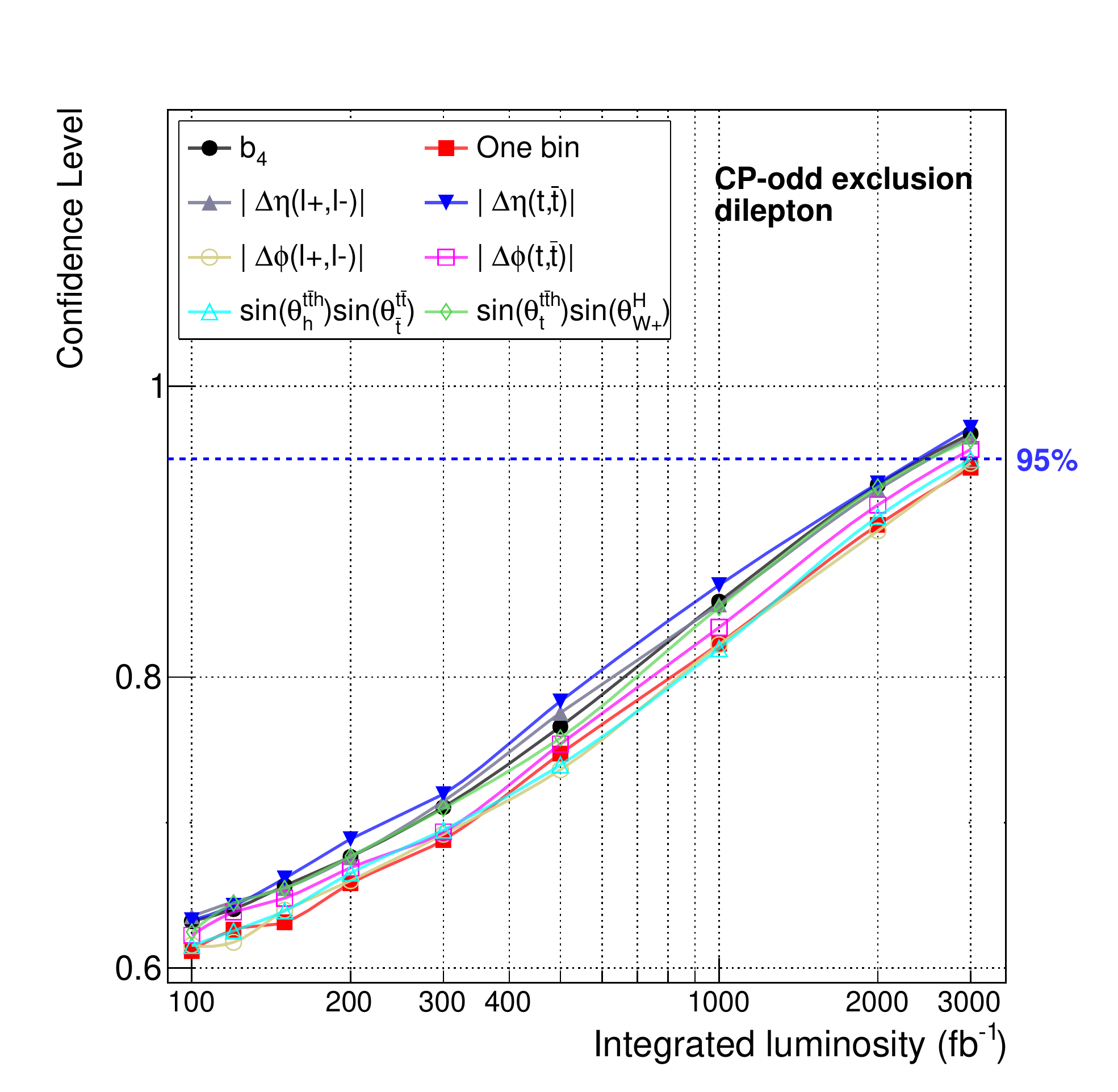}
 \caption{}
 \label{dilep}
\end{subfigure}\\[-2mm]
\begin{subfigure}{0.49\textwidth}
 \includegraphics[width=\textwidth,clip]{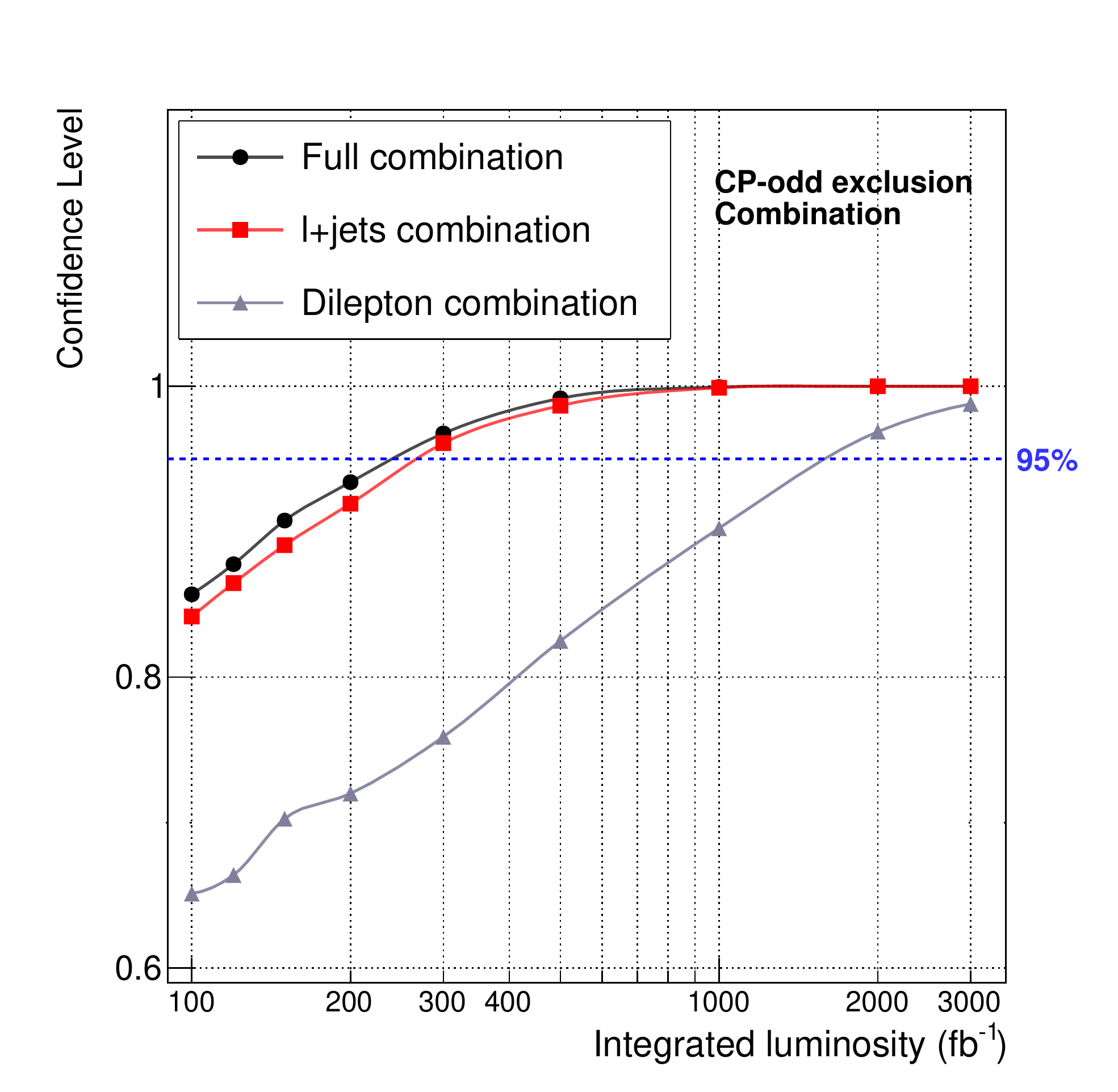}
 \caption{}
 \label{combination}
\end{subfigure}
\hfill
\begin{subfigure}{0.49\textwidth}
 \includegraphics[width=\textwidth,clip]{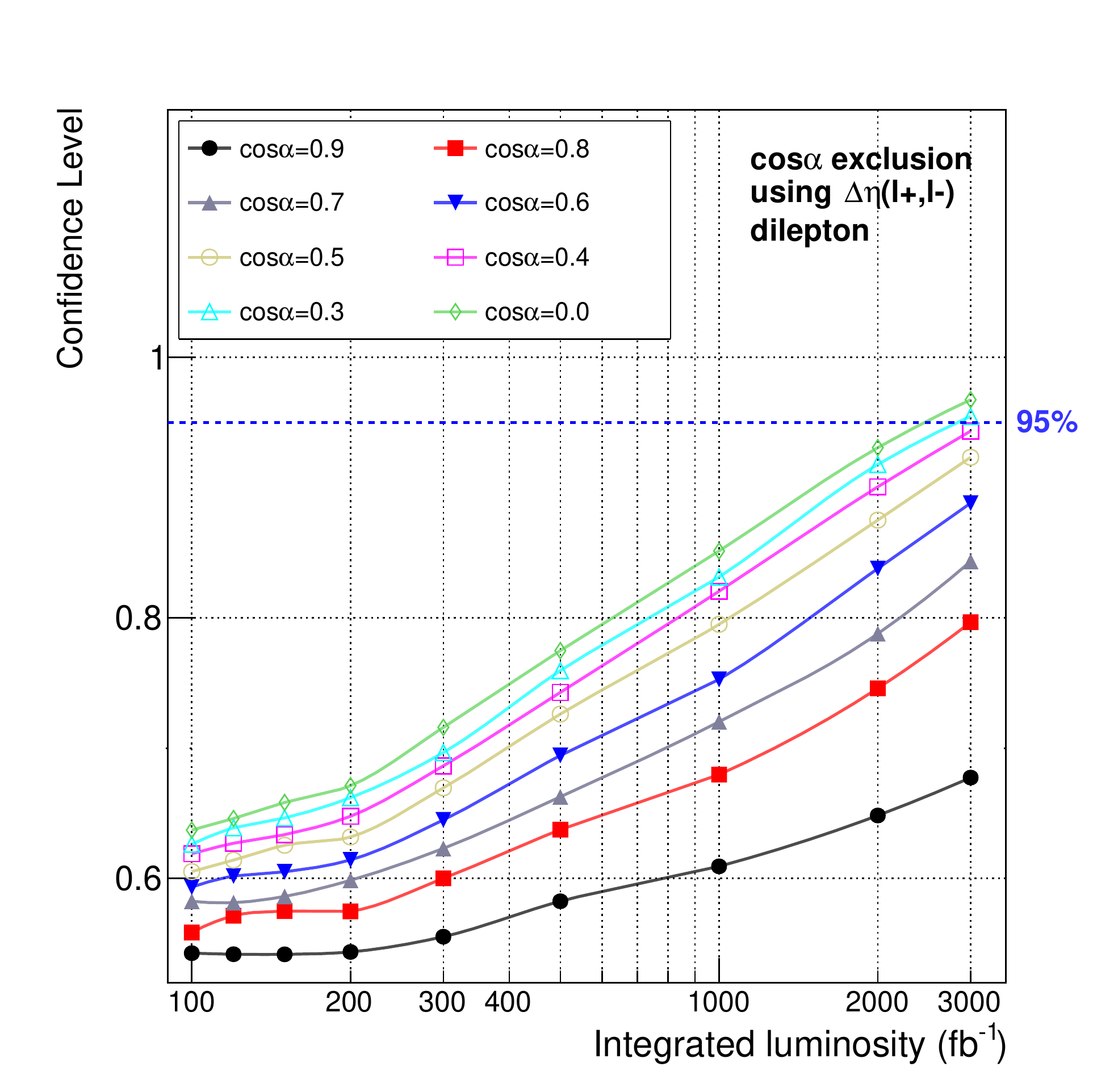}
 \caption{}
 \label{cosaScan}
\end{subfigure}
\caption{Expected CL, assuming the SM, as a function of the integrated luminosity. \ref{semilep} (\ref{dilep}): exclusion of pure CP-odd scenario using the single-lepton (dilepton) selected events, comparing different observables used to extract the CL; \ref{combination}: exclusion of pure CP-odd scenario combining observables in each individual channel and combining both channels (the observables were treated as uncorrelated); \ref{cosaScan}: exclusion of various $\cos\alpha$ values between 0 and 1 with dilepton analysis, using $\Delta\eta(\ell^+,\ell^-)$ as the discriminant observable.}
\end{figure}

Expected confidence levels (CL), assuming the SM holds, for exclusion of the pure CP-odd coupling were computed from likelihood ratios obtained from binned distributions of various discriminant observables. Only statistical uncertainties were considered. Figures \ref{semilep}-\ref{cosaScan} show these CL as a function of integrated luminosity, up to the maximum expected at the HL-LHC (3~ab$^{-1}$). Figure \ref{semilep} (\ref{dilep}) shows the CL obtained using the single-lepton (dilepton) analysis only, comparing different observables used to extract the CL. Figure \ref{combination} shows confidence levels obtained from the combination of different observables in each channel ($b_4$ and $\sin(\theta^{t\bar tH}_{\bar t})\sin(\theta^{H}_{b_H})$ in single-lepton and $\Delta\eta(\ell^+,\ell^-)$, $\Delta\phi(t,\bar t)$ and $\sin(\theta^{t\bar tH}_{t})\sin(\theta^{H}_{W^+})$ in dilepton) and from the combination of the two channels. The observables were treated as uncorrelated. Figure \ref{cosaScan} compares the CL obtained, in the dilepton analysis alone, for the exclusion of various $\cos\alpha$ values, using $\Delta\eta(\ell^+,\ell^-)$ as the discriminant.

The single-lepton channel is much more sensitive than the dilepton, but the combination provides sizeable improvement. The exclusion of a pure CP-odd top quark Yukawa coupling at 95\% CL may be within reach with $\sim 250$ fb$^{-1}$ of LHC data, using only resolved $t\bar t (H\rightarrow b\bar b)$ analyses with leptons. Combination with boosted topologies and other Higgs decay channels could make this achievable with the LHC Run II dataset. The results in figure \ref{cosaScan} suggest that excluding the maximal mixing scenario ($\cos\alpha=\sqrt{2}/2$) requires roughly 3.5 times more luminosity than excluding the pure CP-odd scenario.

\vspace{-7mm}

\Acknowledgements
I acknowlegde funding from Funda\c{c}\~{a}o para a Ci\^{e}ncia e Tecnologia, through grant PD/BD/128231/2016, and from the COST Network CA16201 PARTICLEFACE ``Unraveling new physics at the LHC through the precision frontier''.

\vspace{-6mm}

\bibliographystyle{unsrt}

\begin{thebibliography}{99}
\bibitem{Aaboud:2018urx}
  ATLAS Collaboration,
  Phys.\ Lett.\ B {\bf 784} (2018) 173
  
\bibitem{Sirunyan:2018hoz}
  CMS Collaboration,
  Phys.\ Rev.\ Lett.\  {\bf 120} (2018) no.23,  231801
  
\bibitem{Fontes:2017zfn}
  D.~Fontes {\it et al.}, 
  JHEP {\bf 1802} (2018) 073
  
\bibitem{BarShalom:1995jb}
  S.~Bar-Shalom {\it et al.}, 
  Phys.\ Rev.\ D {\bf 53} (1996) 1162
  [hep-ph/9508314];\\
  D.~Atwood {\it et al.}, 
  Phys.\ Rept.\  {\bf 347} (2001) 1
  [hep-ph/0006032].
  
\bibitem{rohini2}
  F.~Boudjema {\it et al.},
  Phys.\ Rev.\ D {\bf 92} (2015) no.1,  015019
  
\bibitem{demartin}
  F.~Demartin {\it et al.},
  Eur.\ Phys.\ J.\ C {\bf 74} (2014) no.9,  3065
  
\bibitem{cpangvars}
  S.~Amor Dos Santos {\it et al.},
  Phys.\ Rev.\ D {\bf 96} (2017) no.1,  013004
  
\bibitem{gunion}
  J.~F.~Gunion and X.~G.~He,
  Phys.\ Rev.\ Lett.\  {\bf 76} (1996) 4468
  
  
\bibitem{Buckley:2015vsa}
  M.~R.~Buckley and D.~Goncalves,
  Phys.\ Rev.\ Lett.\  {\bf 116} (2016) no.9,  091801
  
\bibitem{duarte}
  D.~Azevedo {\it et al.}, 
  Phys.\ Rev.\ D {\bf 98} (2018) no.3,  033004

\bibitem{madgraph}
  J.~Alwall {\it et al.},
  JHEP {\bf 1407} (2014) 079
  
\bibitem{madspin}
  P.~Artoisenet {\it et al.},
  JHEP {\bf 1303} (2013) 015
  
\bibitem{pythia}
  T.~Sjostrand, S.~Mrenna and P.~Z.~Skands,
  JHEP {\bf 0605} (2006) 026
  
\bibitem{delphes}
  J.~de Favereau {\it et al.} [DELPHES 3 Collaboration],
  JHEP {\bf 1402} (2014) 057

\bibitem{Erdmann:2013rxa}
  J.~Erdmann {\it et al.}, 
   Nucl.\ Instrum.\ Meth.\ A {\bf 748} (2014) 18
  
  

\end{thebibliography}

\end{document}